\documentclass[aip,reprint]{revtex4-1} 	
\usepackage{geometry}              		
\usepackage[dvipdfmx]{graphicx}
\usepackage{xcolor}				
\usepackage[T1]{fontenc}
\usepackage{mathptmx}
\usepackage{float}
\usepackage{amssymb}
\usepackage[utf8]{inputenc}
\usepackage{amsmath}
\usepackage{amsthm}
\usepackage{amsfonts}
\usepackage{bm}
\usepackage{hyperref}
\usepackage[caption=false]{subfig}

\newcommand{\norm}[1]{\left\lVert#1\right\rVert}
\theoremstyle{definition}

\newtheorem{algo}{Algorithm}

\begin{document}

\title{Two Dimensional Swarm Formation in Time-invariant External Potential:\\ Modelling, Analysis, and Control}

\author{Yanran Wang}
\email{y-wang@dove.kuee.kyoto-u.ac.jp}

\author{Takashi Hikihara}%
\email{hikihara.takashi.2n@kyoto-u.ac.jp}
\affiliation{%
 Department of Electrical Engineering, Kyoto University.
}%


\date{\today}
\begin{abstract}
 Clustering formation has been observed in many organisms in Nature. It has the desirable properties for designing energy efficient protocols for Wireless Senor Networks (WSNs). In this paper, we present a new approach for energy efficient WSNs protocol which investigate how cluster formation of sensors response to external time-invariant energy potential. In this approach, the necessity of data transmission to Base Station is eliminated, thereby conserving energy for WSNs. We define swarm formation topology, and estimate the curvature of external potential manifold by analyzing the change of the swarm formation in time. We also introduce a dynamic formation control algorithm for maintaining defined swarm formation topology in external potential.
\end{abstract}

\keywords{Wireless Sensor Networks, nonlinear dynamic, Koopman operator theory, dynamic mode decomposition, curvature, swarm formation}
\maketitle


\section{Introduction}
Wireless Sensor Networks (WSNs) have attracted much attention due to its ability to provide ubiquitous and multi-faceted situational awareness with a host of applications ranging from structural health monitoring, habitat surveillance, and target detection to power system management, smart car parking, and wireless luggage tags \cite{RN2, RN3, RN4, RN1, RN5}. WSN depends on spatially distributed sensor node to measure and collect the desire environmental data within its sensing range, and then transmit to a control center called Base Station (BS). The ideal WSN should be autonomous, robust, scalable, and with extended network lifetime. However, WSNs are usually deployed in hostile environments where energy resources is limited. As energy is constrained and data transmission is most energy costly, WSNs algorithms need to be architect in ways where data transmission, especially to BS, is minimized.
Clustering become an often utilized technique in designing energy efficient algorithm for WSNs.

Cluster formations can be readily observed in Nature, such as bird flocking and fish schooling. These organism systems are expressing collective motion behaviours and are studied in a relatively novel interdisciplinary field of research, Swarm Intelligence (SI). Individual agents in the swarm are simple agents with limited sensing abilities and computational rules, and interacting with each other locally. Nevertheless, the swarm as a whole demostrates emergent global behaviours which are unknown to the individuals. SI systems have the desirable properties of being distributed, autonomous, scalable, and robust \cite{RN6}. All of which are key in designing algorithms for WSNs.

Plentiful of clustering algorithm has been developed \cite{8908057, RN7, RN8}. However, besides the main objective of energy conservation, the existing clustering algorithms mainly focus on the optimalization of sensor protocal routines to enhance WSNs in scalability, fault-tolerance, data aggregation, load balancing and network topology stability. All of which data transmission to BS is unavoidable. To address the energy efficiency challenge from a new perspective, this paper combines SI concept with WSNs. We focuses on designing WSNs algorithm where desired environmental information are obtained through analyzing the change in sensor cluster formations (swarm formation) rather than information collected directly through individual sensors, thereby minimizing the energy expenditure in data tranmission between sensors and eliminates the necessity of BS.

We identify the envrionmental information as a external potential hypersuface $M$ of $(n-1)$-dimensions ($n=3$ or $n=4$). Mathematically, $M$ is a Riemannian manifold defined by the set of solutions to a single equation
\begin{align}
 F(x_1, \ldots, x_n)=0,
\end{align}
where $F$ is a $C^{\infty}$ function.
We introduce a formation analysis algorithm which uses swarm formation in external potential to estimate curvature, which is invariant under isometry, of the manifold $M$.

This paper is organized as the follows. In section \ref{Swarm formation}, we introduce the topology of swarm lattice formation and demonstrate how to construct such formation in an arbitary external potential. In section \ref{Formation analysis}, we explain the theoretical basis and computations for the formation analysis algorithm. In section \ref{Formation control}, we formulate an formation control algorithm which shows how the swarm lattice formation can be obtained by dynamical control of the individual agents. In section \ref{Simulation and discussion}, we discuss the simulation results of both formation analysis and formation control algorithms.

\section{Swarm formation}
\label{Swarm formation}
The topology definition of the swarm formation is inspired by Reynolds \cite{RN13} and Olfati-Saber \cite{RN12}. In this paper, we consider line formation. In the line formation, agents are being divided into two groups: one head agent and the following agents. The head agent acts as an initial stimuli to the swarm motion. Its dynamics can be predefined and are not affected by the following agents in the swarm.

Define a $\textit{path graph}$ $P_n$ as a pair $(\mathcal{V},\mathcal{E})$ that consists of a set of vertices $\mathcal{V} = \{v_1,v_2, \dots, v_n\}$ and a set of edges $\mathcal{E}$ such that $\mathcal{E} \subseteq \{v_i, v_{i+1}\}$, where $i=1, 2, \dots, n-1$. Each vertex represents an agent of the swarm, while the edges represent the inter-agent communications. Let $q_i \in \mathbb R$ denote the position of agent $v_i$ for all $v_i \in \mathcal{V}$. The vector $q=(q_1,...,q_n)^T$ is the configuration of all agents of the swarm. The inter-agent distance, that is the length of edges, is defined to be the geodesic length between two connected vertices over $G(q)$. To maintain identical inter-agent distance, we consider an algebraic constraint on the edges,
\begin{align}
 dis(\mathcal{E}_{v_i,v_{i+1}}) = d, \hspace{5mm} \forall \,\,v_i \in V, \hspace{5mm} d \in \mathbb R.
 \label{latticeconstraint}
\end{align}
A configuration $q$ that satisfying the set of constraints in (\ref{latticeconstraint}) is refered as a \textit{lattice formation}.

To formulate lattice formation in any arbitary external potential $M$, we need to first define the trajectory of the head agent, then construct a representation of edges -- a parallel vector field that is metrically orthogonal to the head agent trajectory, and finally calculate the geodesic deviation vector field. The trajectories of the following agents are the integral curvces of the geodesic deviation vector field. The theoretical basis \cite{RN11} and details of the formulation are discussed as follows.

Let $U \subseteq \mathbb{R}^n$ be a non-empty open subset and $F: U \rightarrow \mathbb{R}$ a $C^{\infty}$ function defining the external potential. Let $M \subseteq U \times \mathbb{R}$ be the
graph of $f$. The closed subset $M$ in $U \times \mathbb{R}$ projects homeomorphically onto $U$ with inverse
$(x_1, \ldots, x_n) \mapsto (x_1, \ldots, x_n, F(x_1, \ldots, x_n))$
that is a smooth mapping from $U$ to $U \times \mathbb{R}$. $M$ is a closed smooth submanifold of $U \times \mathbb{R}$. Using the standard Riemannian metric on $U \times \mathbb{R} \subseteq \mathbb{R}^{n+1}$, the induced metric $g$ on $M$ at a point $p \in M$ is
\begin{align}
 g(p) = \langle \partial_{q_i}|_p, \partial_{q_j}|_p\rangle_p \, dq_i(p) \otimes dq_j(p)
\end{align}
with coordinate chart $\{q_i\}$ on $M$. Each $\partial_{q_i}|_p \in T_pM$ can be represented as a linear combination of $\{\partial_{x_i}|_p\} \in T_p(\mathbb{R}^{n+1})$, given as
\begin{align}
 \partial_{q_i}|_p = \partial_{x_i}|_p + \partial_{x_i}f(p)\partial_{x_{n+1}}|_p.
\end{align}
Consider the aforementioned graph $M$ as a $C^{\infty}$ Riemannian manifold. Given a curve, $C:[ a, b ] \longrightarrow M$, a \textit{vector field} $X$ along $C$ is any section of the tangent bundle $TM$ over $C$ ($X:[a,b]\longrightarrow TM$, projection $\pi: TM \longrightarrow M$, such that $\pi \circ X = C$). If $M$ is a smooth manifold, all vector field on the manifold are also smooth. We denote the collection of all smooth vector fields on manifold $M$ as $\mathfrak{X}(M)$. For a Riemannian manifold $(M, g)$, the \textit{Levi-Civita connection} $\nabla_g$ on M is the unique connection on $TM$ that has both metric compatibility and torsion freeness. The Christoffel symbols of the second kind are the connection coefficients (in a local chart) of the Levi-Civita connection denoted as
\begin{align}
 \Gamma^a_{\,\,bc}=\frac{1}{2}g^{a d}(\partial_cg_{db}+\partial_bg_{dc}-\partial_dg_{bc}).
\end{align}
For a Riemannian manifold $(M, g)$, a curve is called \textit{geodesic} with respect to the connection $\nabla_g$ if its acceleration it zero. That is a curve $\gamma$ where $\nabla_{\dot\gamma}\dot\gamma = 0$. A geodesic curve in n-dimensional Riemannian manifold can be expressed as a system of second order ordinary differential equations,
\begin{align}
 \frac{d^2\gamma^\lambda}{dt^2}+\Gamma^\lambda_{\,\,\mu \nu}\frac{d\gamma^\mu}{dt}\frac{d\gamma^\nu}{dt}=0.
 \label{geodesic formula}
\end{align}
All geodesics are the shortest path between any two points on the manifold.

We predefine the trajectory of the head agent to be a geodesic curve on manifold. Base on (\ref{geodesic formula}), for a two-dimensional manifold $(M, g)$, head agent trajectory $h(t)$ can be expressed as a system of ordinary differential equations
\begin{align}
 \begin{split}
  \dot h_1 &= h_3,\\
  \dot h_2 &= h_4,\\
  \dot h_3 &= -\Gamma^x_{\,\,xx}(h_3)^2-2\Gamma^x_{\,\,xy}h_3h_4-\Gamma^x_{\,\,yy}(h_4)^2,\\
  \dot h_4 &=-\Gamma^y_{\,\,xx}(h_3)^2-2\Gamma^y_{\,\,xy}h_3h_4-\Gamma^y_{\,\,yy}(h_4)^2,
 \end{split}
 \label{head_curve_2d}
\end{align}
where basis $\{x,y\}$ are used in the index, and $\dot h$ are the first derivative with respect to time. Each Christoffel symbols depends entirely on the metric at a certain point in $M$ in terms of basis $\{x,y\}$. Given initial conditions, $[x_1, x_2, \dot x_1, \dot x_2 ]$, (\ref{head_curve_2d}) is guaranteed to have a solution according to the Picard-Lindel$\rm{\ddot o}$f theorem. This choice of head agent trajectory is made to simplify agent's dynamic control. On the geodesic, the head agent is traveling with constant velocity given initial position and velocity, thus no requirement for outside reference beacon.

The distance of the edges of lattice formation needs to be identical. For edges representation, a parallel vector field $K$ that is orthogonal to the head agent trajectory $h(t)$ is constructed as
\begin{align}
 \nabla_K K=0.
\end{align}
$K$ forms a family of geodesics where we require the initial position and velocity conditions $(u^j_0, v^j_0)$ for $j$th geodesic $\gamma_j \in K$  are $u^j_0=h(t_j)$, $v^j_0=v: g(v, \dot{h}(t_j))=0$, $j \in \mathbb{Z}$. The frequency of inter-agent communications is defined by  the number of geodesics in $K$ within given traveling time.

\begin{figure}[h]
 \includegraphics[width=1\linewidth]{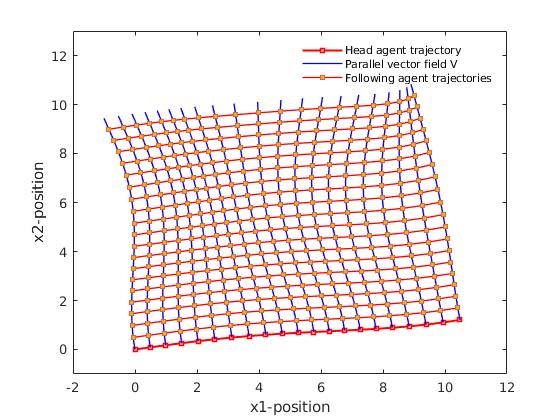}
 \caption{Two dimensional visual example of swarm lattice formation in external potential $F=\sin (\frac{x_1}{a}) + \cos (\frac{x_2}{a})$, $a=2$. Travel time $t=20$. Head agent in red, with initial condition $[0,\, 0, \, \cos (\pi/18), \, \sin(\pi/18)]$; following agents in orange.}
 \label{2dt_case4_inew_a2_t1}
\end{figure}

The separation vector $s(t)$ connects a point $\gamma (t)$ on one geodesic to a point $\gamma(t)+s(t)$ on a nearby geodesic at the same time. For parallel vector field $K$, we can  construct a separation vector field $S$ such that $s \in S$ are the separation vector discribed above. For the swarm lattice formation, the following agents' trajectories are the integral curves of $S$.  Fig. \ref{2dt_case4_inew_a2_t1} gives a visual example of two-dimensinal swarm lattice formation traveling in external potential.

\section{Formation analysis}
\label{Formation analysis}
In Fig. \ref{2dt_case4_inew_a2_t1}, one can notice the change in swarm lattice formation as it travels due to the curvature of the external potential manifold. As an invariant property under isometry, curvature tensor gives valuable information about the external potential manifold itself. We quantify this change as the acceleration of the separation vectors $s \in S$ along $K$, which is equivalent to the change in the difference of neighbouring agents' velocities. Geodesic deviation equation relates the acceleration of the separation vector between two neighbouring geodesic curves to Riemann curvature tensor. The theoretical elements \cite{RN11,RN10} and analysis method are discribed in details as follows.

The Riemann curvature tensor is a $(1, 3)$ tensor defined through the Lie bracket on $(M, g)$ as
\begin{align}
 R(X, Y)Z =\nabla_{[X,Y]} Z - \nabla_X\nabla_Y Z + \nabla_Y\nabla_X Z,
\end{align}
where $X, Y, Z \in \mathfrak{X}(M)$, and $R(X, Y)Z$ is vector-valued. $R(X, Y)Z$ can be express in local chart as
\begin{align}
 R^{\rho }_{\,\,\,\sigma \mu \nu }=\partial _{\mu }\Gamma ^{\rho }_{\,\,\nu \sigma }-\partial _{\nu }\Gamma ^{\rho }_{\,\,\mu \sigma }+\Gamma ^{\rho }_{\,\, \mu \lambda }\Gamma ^{\lambda}_{\,\,\nu \sigma}-\Gamma ^{\rho}_{\,\,\nu \lambda}\Gamma ^{\lambda }_{\,\,\mu \sigma }.
\end{align}

By our definition in section \ref{Swarm formation}, the separation vector $s$ can be written as vectors,
\begin{align}
 s(t)=\widetilde{\gamma}(t)-\gamma(t),
 \label{deviation vector}
\end{align}
between two nearby geodesics. The separation acceleration field $W$ is
\begin{align}
 W=\nabla_{\dot\gamma}\nabla_{\dot\gamma} S = \nabla_{\dot\gamma}V,
\end{align}
where $V$ is the separation velocity field . In a local chart, $W$ and $V$ can be expressed as
\begin{align}
 V^\rho & =\frac{dS^\rho}{dt}+\Gamma^\rho_{\mu \nu}\dot\gamma^\mu S^\nu,
 \label{sep_vector_eq1}
 \\
 W^\rho & =\frac{dV^\rho}{dt}+\Gamma^\rho_{\lambda \sigma}\dot\gamma^\lambda V^\sigma,
 \label{sep_vector_eq2}
\end{align}
where $\dot\gamma \equiv \displaystyle\frac{d\gamma}{dt}$.
Combining (\ref{sep_vector_eq1}) and (\ref{sep_vector_eq2}) gives
\begin{align}
 \begin{split}
  W^\rho=&\frac{d^2S^\rho}{dt^2}+2\Gamma^\rho_{\,\,\mu \nu}\dot\gamma^\mu\frac{dS^\nu}{dt}+\frac{\partial\Gamma^\rho_{\,\,\mu \nu}}{\partial\gamma^\lambda}\dot\gamma^\lambda\dot\gamma^\mu S^\nu\\
  &+\Gamma^\rho_{\,\,\mu \nu}\ddot\gamma^\mu S^\nu+\Gamma^\rho_{\,\,\lambda \sigma}\Gamma^\sigma_{\,\,\mu \nu}\dot\gamma^\mu \dot\gamma^\lambda S^\nu,
 \end{split}
 \label{W}
\end{align}
where $\ddot\gamma \equiv \displaystyle\frac{d^2\gamma}{dt^2}$. Rearranging (\ref{deviation vector}) to be
\begin{align}
 \widetilde\gamma(t)=\gamma(t)+s(t),
\end{align}
where separation vector $s(t)$ is treated as an expansion parameter. Make use of the fact that both $\gamma$ and $\widetilde \gamma$ are geodesics, then we have
\begin{align}
 0=\frac{d^2S^\rho}{dt^2}+2\Gamma^\rho_{\,\,\mu \nu}\dot\gamma^\mu\frac{dS^\nu}{dt}+\frac{\partial\Gamma^\rho_{\,\,\mu \nu}}{\partial\gamma^\lambda}\dot\gamma^\mu\dot\gamma^\nu S^\lambda.
 \label{sep2}
\end{align}
Inserting (\ref{sep2}) into (\ref{W}), the first-order geodesic deviation equation is
\begin{align}
 \begin{split}
  W^\rho=&-(\frac{\partial\Gamma^\rho_{\,\,\mu \lambda}}{\partial\gamma^\nu}-\frac{\partial\Gamma^\rho_{\,\,\mu \nu}}{\partial\gamma^\lambda}+\Gamma^\sigma_{\,\,\mu \lambda}\Gamma^\rho_{\,\,\sigma \nu}-\Gamma^\rho_{\,\,\lambda \sigma}\Gamma^\sigma_{\,\,\mu \nu})\\
  &\dot\gamma^\mu S^\nu\dot\gamma^\lambda\\
  \equiv& -R^\rho_{\,\,\,\mu \nu \lambda}\dot\gamma^\mu S^\nu \dot\gamma^\lambda.
 \end{split}
 \label{Wfinal}
\end{align}
A vector field $J$ along a geodesic $\gamma$ is called a Jacobi field if
\begin{align}
 \ddot J+R(J, \dot\gamma)\dot\gamma = 0
\end{align}
where $\ddot J \equiv \nabla_{\frac{d}{dt}}\nabla_{\frac{d}{dt}}J$, and $\dot\gamma\equiv\frac{d\gamma}{dt}$. Obviously, the deviation vector field $S$ is a Jacobi field.

Furthermore, sectional curvature is a equivalent but more geometrical description of the curvature of Riemannian manifolds. Let tangent 2-plane, $\Pi_p$, be the two dimensional subspace in $T_{p}M$ defined as $\Pi_p \equiv span \{u, v\}$, with $u, v \in T_{p}M$.  Sectional curvature $\kappa$ of $(M,g)$ at a point $p \in M$ with respect to the plane $\Pi_p$ is defined as
\begin{align}
 \begin{split}
  \kappa(\Pi_p)&=\kappa(X_p, Y_p) \\
  &=\frac{\langle R(X,Y)X,Y\rangle_p}{|X|^2_p|Y|^2_p-\langle X,Y\rangle^2_p}
 \end{split}
 \label{sectional curvature}
\end{align}
Substitude the vector fields $K$ and $S$ we constructed in section \ref{Swarm formation} into (\ref{sectional curvature}), we have
\begin{align}
 \begin{split}
  \kappa(K, S)&=\frac{\langle R(K, S)K,S\rangle}{|K|^2|S|^2-\langle K,S\rangle^2}.
 \end{split}
\end{align}
Also $S$ is in fact a Jacobi field along $K$ that is also orthogonal to $K$. That is $W \equiv \nabla^2_\tau S = -R(K, S)K$, and $\langle K,S\rangle = 0$. Combining with the fact that the integral curves of $K$ are geodesics set to have velocity $|\dot\gamma| = 1$, the equation for sectional curvature is simplified to
\begin{align}
 W=-\kappa(K, S)S.
 \label{Wfinalfinal}
\end{align}
For a two dimensional Riemannian manifold $(M, g)$, there is only one sectional curvature at each point $p \in M$.

In summary, by recording the change of swarm lattice formation in external potential, observer can quantify two variables: $S$, agents' velocities; $W$, the second-order differences of nearby agents' velocities. The geodesic deviation equation (\ref{Wfinalfinal}) relates $W$ and $S$ by sectional curvature $\kappa$ of the external potential manifold.

\section{Formation control}
\label{Formation control}
To stay in lattice formation as the swarm travling through external potential, individual agent needs dynamic control of its trajectory. The trajectory of individual agent can be considered as a dynamical system of the form \cite{RN15}
\begin{align}
 \frac{d}{dt}\mathbf{x}(t)=\mathbf{F}(\mathbf{x}(t)),
 \label{dynamic system}
\end{align}
where $\mathbf{x}$ is the state of the system, that is the position of the agent, and $\mathbf{F}$ is a vector field depends on the system. In practice, we consider the equivalent discrete-time dynamical system,
\begin{align}
 \mathbf{x}_{k+1} = \mathbf{F}(\mathbf{x}_k),
\end{align}
where $\mathbf{x}_k$ is the sampling of agent trajectory in (\ref{dynamic system}) discretely in timesteps $\Delta t$. If we can assume the system is of linear dynamic, then we can work with the form
\begin{align}
 \frac{d}{dt}\mathbf{x}=\mathbf{Ax},
\end{align}
where it admits a closed-form solution
\begin{align}
 \mathbf{x}(t_0+t)=e^{\mathbf{A}t}\mathbf{x}(t_0).
\end{align}
The entire system dynamic is characterized by the eigenvalues and eigenvectors of the matrix $\mathbf{A}$. Using spectral decomposition, one can simplify the dynamic system to
\begin{align}
 \frac{d}{dt}\mathbf{w}=\mathbf{\lambda w},
\end{align}
with $\mathbf{A}=\mathbf{P\lambda P}^{-1}$, and $\mathbf{w}=\mathbf{P}^{-1}\mathbf{x}$.
Using $\mathbf{A}$, we can also predict agents' trajectories in time, thereby controlling the swarm to be in its lattice formation as time evolves.

Even though it is desirable to work with linear dynamical systems, the curvature property of the external potential manifold, and thus the trajectories of agents are essentially nonlinear. To analyze nonlinear dynamic with linear technique, we utilize Koopman operator theory.

\subsection{Koopman operator theory and dynamic mode decomposition}
Bernard O. Koopman has proved the posibility of representing a nonlinear dynamical system in terms of an infinite-dimensional linear operator acting on a Hilbert space of measurement functions of the state of the system. The basic elements of Koopman spectral analysis is dicussed below \cite{Koopman, book,Yoshihiko,7403328}.

Consider a real-valued measurement function $g: \mathbf{M} \longrightarrow \mathbb{R}$, known as \textit{observables}, which belongs to the infinite dimensional Hilbert space. The Koopman operator $K_t$ is an infinite-dimensional linear operator that acts on the observable $g$ as
\begin{align}
 K_tg=g \circ \mathbf{F}_t,
\end{align}
where $\mathbf{F}_t$ is the system dynamic, and $\circ$ is the composition operator.
For discrete-time system with timestep $\Delta t$, it becomes
\begin{align}
 g(\mathbf{x}_{k+1})= K_{\Delta t}g(\mathbf{x}_k).
\end{align}
Even though Koopman operator is linear, it is also infinite dimensional. Thus it is important to identify key measurement functions that evolve linearly with the flow of the dynamic. Eigen-decomposition of the Koopman operator can provide us with such a set of measurement functions that captures the dynamics of the system and also behaves linearly in time. A discrete-time Koopman eigenfunction $\varphi(\mathbf{x})$ and its corresponding eigenvalue $\lambda$ satisfies
\begin{align}
 \varphi(\mathbf{x}_{k+1})=K_{\Delta t}\varphi(\mathbf{x}_k)=\lambda\varphi(\mathbf{x}_k).
\end{align}
Nonlinear dynamics become linear in these eigenfunction coordinate.

In a general dynamic system, the measurement functions can be arranged into a vector $\mathbf{g}$:
\begin{align}
 \mathbf{g}(\mathbf{x})=\begin{bmatrix}
  g_1(\mathbf{x}) \\
  g_2(\mathbf{x}) \\
  \vdots          \\
  g_m(\mathbf{x}) \\
 \end{bmatrix}.
\end{align}
Each measurement functions may be expanded in terms of eigenfunctions $\varphi_j(\mathbf{x})$, thus vector $\mathbf{g}$ can be written as:
\begin{align}
 \mathbf{g}(\mathbf{x})=\sum_{j=1}^{\infty} \varphi_j(\mathbf{x})\mathbf{v}_j,
\end{align}
where $\mathbf{v}_j$ is the $j$-th Koopman mode associated with the eigenfunction $\varphi_j$.
Given this decomposition, we can represent the dynamics of the system in terms of measurement function $\mathbf{g}$ as
\begin{align}
 \begin{split}
  \mathbf{g}(\mathbf{x}_k)&=K^k_{\Delta t}\mathbf{g}(\mathbf{x}_0)  \\
  &=K^k_{\Delta t}\sum_{j=0}^{\infty} \varphi_j(\mathbf{x}_0)\mathbf{v}_j \\
  & =\sum_{j=0}^{\infty} K^k_{\Delta t}\varphi_j(\mathbf{x}_0)\mathbf{v}_j \\
  & =\sum_{j=0}^{\infty} \lambda^k_j\varphi_j(\mathbf{x}_0)\mathbf{v}_j.
 \end{split}
\end{align}
The sequence of triples $\{(\lambda_j, \varphi_j, \mathbf{v}_j )\}^{\infty}_{j=0}$ is the Koopman mode decomposition.

Finding such Koopman mode is extremely diffcult even for system with known governing equations. For systems with unknown governing equation, such as in our situation, dynamic mode decomposition (DMD) algorithm is adopted. As one of the modal decomposition technique, DMD is first introduced in fluid dynamics community to analyze high-dimensional fluid state. It provides information about the dynamics of a flow in superposition of modes (similar to eigenmodes), even if the flow is nonlinear. DMD analysis can be considered as a approximation to Koopman spectral analysis, and also provides a numerial method for computing Koopman modes. In this paper, the formation control algorithm is inspired by online DMD framework from Zhang et al. \cite{OnlineDMD}.

We consider a sequential set of data vectors $\{\mathbf{x}_0, \mathbf{x}_1, \dots, \mathbf{x}_m\}$, where each $\mathbf{x}_k \in \mathbb{R}^n$, are system states of time $t_0$ to $t_m$. These data can be arraged into two matrices
\begin{align}
 \mathbf{X} & =\begin{bmatrix}
  \vert        & \vert        &       & \vert            \\
  \mathbf{x}_0 & \mathbf{x}_1 & \dots & \mathbf{x}_{m-1} \\
  \vert        & \vert        &       & \vert            \\
 \end{bmatrix},
 \\
 \mathbf{Y} & =\begin{bmatrix}
  \vert        & \vert        &       & \vert          \\
  \mathbf{x}_1 & \mathbf{x}_2 & \dots & \mathbf{x}_{m} \\
  \vert        & \vert        &       & \vert          \\
 \end{bmatrix}.
\end{align}
We assume there exist an operator $\mathbf{A}$ that approximates the nonlinear dynamic of the system as
\begin{align}
 \mathbf{x}_{k+1}\approx \mathbf{A}\mathbf{x}_k.
\end{align}
Then the best-fit operator $\mathbf{A}$ is defined as
\begin{align}
 \mathbf{A} = \mathop{\arg\min}_{\mathbf{A}} \norm{\mathbf{Y}-\mathbf{AX}}_F,
\end{align}
where $\norm{\cdot}_F$ is the Frobenius norm.
The unique solution to least-square problem is given by
\begin{align}
 \mathbf{A}=\mathbf{YX}^{\dagger},
 \label{matrix a}
\end{align}
where $X^{\dagger}$ denotes the Moore-Penrose pseudoinverse of $\mathbf{X}$.

Having $\mathbf{A}$, computed from data from $t_0$ up to $t_m$, we can predict $x_{m+1}$ of the system, that is controlling agent's trajectory. Since new data are feeded in to the system as time progresses, $\mathbf{A}$ is also changing. Unlike the standard DMD algorithm, online DMD algorithm updates operator $\mathbf{A}$ using the new system data, providing a more reliable operator $\mathbf{A}$ for the prediction of future system states. The algorithm does not compute the least-square problem of the whole system once new data is been updated. Instead, it computes $\mathbf{A}_{k+1}$ given $\mathbf{A}_k$ and new pairs of data $(x_{k+1},y_{k+1})$, on the assumption that $\mathbf{A}_{k+1}$ is close to $\mathbf{A}_k$ in some sense.

Using (\ref{matrix a}), we have
\begin{align}
 \mathbf{A}_k=\mathbf{Y}_k\mathbf{X}_k^T(\mathbf{X}_k\mathbf{X}_k^T)^{-1}.
\end{align}
We define two new matrices $\mathbf{P}_k$ and $\mathbf{Q}_k$,
\begin{align}
 \mathbf{Q}_k & =\mathbf{Y}_k\mathbf{X}_k^T,        \\
 \mathbf{P}_k & =(\mathbf{X}_k\mathbf{X}_k^T)^{-1},
 \label{p and q}
\end{align}
so that $\mathbf{A}_k = \mathbf{Q}_k \mathbf{P}_k$. $\mathbf{P}_k$ is well-defined if we ensure $ \mathbf{X}_k\mathbf{X}_k^T$ is invertible.
The operator $\mathbf{A}$ at time $t_{k+1}$ is related to $\mathbf{A}_k$ as
\begin{align}
 \begin{split}
  \mathbf{A}_{k+1}&=\mathbf{Q}_{k+1}\mathbf{P}_{k+1}\\ &=(\mathbf{Q}_{k}+\mathbf{y}_{k+1}\mathbf{x}_{k+1}^T)(\mathbf{P}_{k}^{-1}+\mathbf{x}_{k+1}\mathbf{x}_{k+1}^T)^{-1}.
 \end{split}
 \label{A k+1 matrix}
\end{align}
Using Sherman-Morrison formula, we can express $\mathbf{P}_{k+1}$ as
\begin{align}
 \begin{split}
  \mathbf{P}_{k+1}&=(\mathbf{P}_{k}^{-1}+\mathbf{x}_{k+1}\mathbf{x}_{k+1}^T)^{-1}\\
  &=\mathbf{P}_{k}-\frac{\mathbf{P}_k\mathbf{x}_{k+1}\mathbf{x}_{k+1}^T\mathbf{P}_k}{1+\mathbf{x}_{k+1}^T\mathbf{P}_k\mathbf{x}_{k+1}}.
 \end{split}
 \label{P matrix}
\end{align}
$\mathbf{P}_{k+1}$ can be updated more efficiently, without the computation of inverses.
Combining (\ref{P matrix}) and (\ref{A k+1 matrix}), we obtain the formula
\begin{align}
 \mathbf{A}_{k+1}=\mathbf{A}_{k}+\frac{(\mathbf{y}_{k+1}-\mathbf{A}_{k}\mathbf{x}_{k+1})\mathbf{x}_{k+1}^T\mathbf{P}_{k}}{1+\mathbf{x}_{k+1}^T\mathbf{P}_k\mathbf{x}_{k+1}}.
\end{align}
$\mathbf{A}_{k+1}$ is computed using $\mathbf{A}_k$ and new data pair $\{\mathbf{x}_{k+1},\mathbf{y}_{k+1}\}$.

\subsection{Formation control algorithm}
\begin{figure}[h]
 \centering
 \includegraphics[width=1\linewidth]{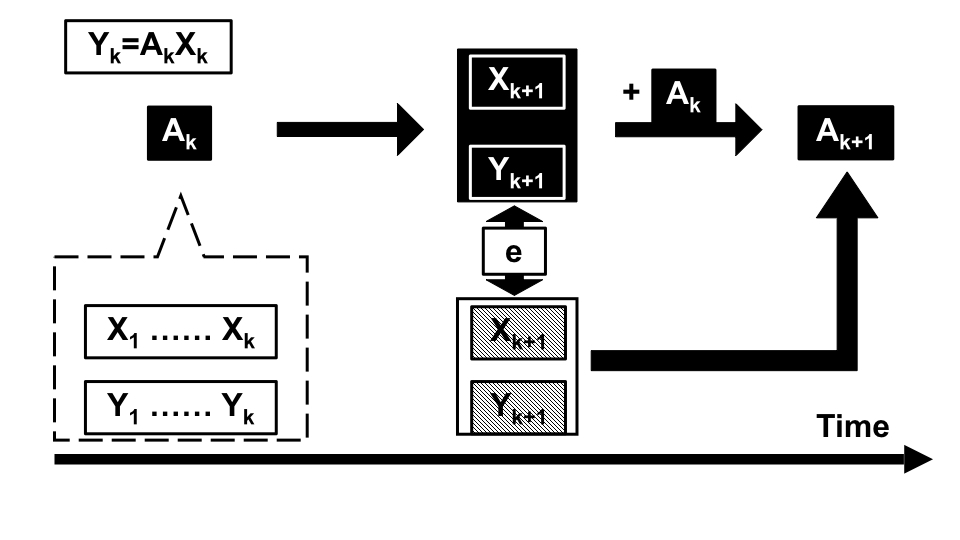}
 \caption{Dynamic control algorithm scheme. White box indicates raw system data obtained from measurements; black box indicates algorithm calculated data; strip box indicates corrected data. }
 \label{dynamic_control_scheme}
\end{figure}

The aforementioned Online-DMD algorithm is a data-driven algorithm which predicts nonlinear dynamics. Therefore, even without knowledge of the external energy potential, it is still possible to control the lattice formation of the swarm. The general scheme of the control algorithm is shown in Fig. \ref{dynamic_control_scheme}.
Notice, the predicted agent trajectory is not equivalent to the ideal trajectory of lattic formation, correction term need to be added to the prediction. The details of Online DMD algorithm modeled specific to control the lattice formation of swarm is shown next.
\begin{algo}
 (Online DMD)
 \begin{enumerate}
  \item Collecting system data as it evolves in time. Arrange data into two matrices
        \begin{align*}
         \mathbf{X} & \equiv [x_0\,\, x_1\,\,\, \dots \,\,\, x_{k-1} ], \\
         \mathbf{Y} & \equiv [x_1\,\, x_2\,\,\, \dots \,\,\, x_k].
        \end{align*}
  \item Compute $\mathbf{A}_k$ and $\mathbf{P}_k$ from (\ref{matrix a}) and (\ref{p and q}).
  \item Predict $\mathbf{y}_{k+1}$ from $\mathbf{y}_{k+1}=\mathbf{A}_k\mathbf{x}_{k+1} = \mathbf{A}_k\mathbf{y}_k$.
  \item Correcting $\mathbf{y}_{k+1}$ by agent's measurements.
  \item Update $\mathbf{A}_k$ and $\mathbf{P}_k$ using corrected data pair $(\mathbf{x}_{k+1},\mathbf{y}_{k+1})$, according to (\ref{A k+1 matrix}) and (\ref{P matrix}).
 \end{enumerate}
\end{algo}
This algorithm is scalable to the rest of the following agents in the swarm, thereby the whole swarm can be dynamically controlled to stay in lattice formation.

\section{Simulation and discussion}
\label{Simulation and discussion}
\subsection{Formation analysis algorithm}
The formation analysis algorithm in section \ref{Formation analysis} is perfomred on swarm lattice formation in three different external potentials:
\begin{enumerate}
  \item elliptic paraboloid: $\displaystyle\frac{x_1^2+x_2^2}{a}$,
  \item hyperbolic paraboloid: $\displaystyle\frac{x_1^2-x_2^2}{a}$,
  \item sinusoidal and cosusoidal: $ \sin(\displaystyle\frac{x_1}{a}) + \cos(\displaystyle\frac{x_2}{a})$,
\end{enumerate}
parametrized by $a$. All simulations used 100 following agents, with traveling time $t=10$.

\begin{figure}[ht]
 \includegraphics[width=1\linewidth]{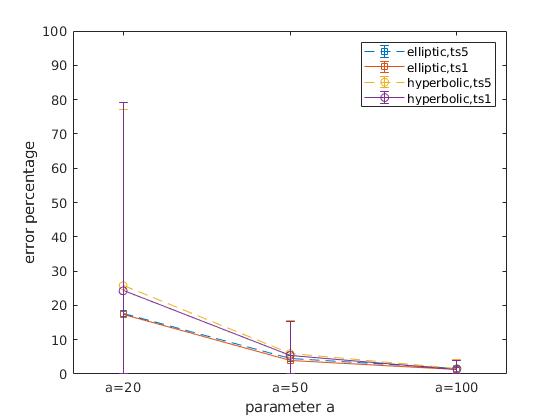}
 \caption{Percentage error for sectional curvature estimation. All simulations used same initial conditions for head agent and have inter-agent distance constraint $d=0.1$; communication frequency $t_{s1} = 0.1, t_{s2}=0.5$. Square represents the mean error.}
 \label{curva_errorbar_case2and3_a}
\end{figure}
\begin{figure}[!h]
 \includegraphics[width=1\linewidth]{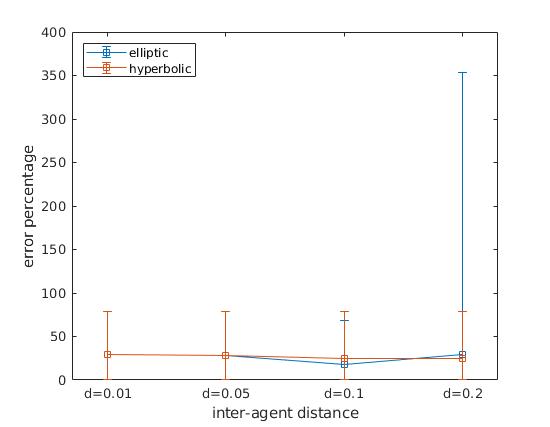}
 \caption{Percentage error for sectional curvature estimation. All simulations used same initial conditions for head agent and communication frequency $t_{s} = 0.1$. Square is the mean error.\vspace{0.5cm}}
 \label{curva_errorbar_case2and3_d}
\end{figure}
The algorithm is able to estimate the sectional curvature of all three external potential manifolds with some extend of accuracy. Overall, the estimation accuracy increase with decrease of curvature; while frequency of inter-agent communications and inter-agent distance constraint does not significantly affect algorithm accuracy shown in Fig. \ref{curva_errorbar_case2and3_a} and Fig. \ref{curva_errorbar_case2and3_d}. However, inter-agent distance directly affect the area of which the swarm is sensing. The accuracy of the algorithm is linked to the following features of the sensing area. For external potential $F(x_1, x_2)= \sin(\frac{x_1}{a}) + \cos(\frac{x_2}{a})$, both features is repsented depending on the swarm sensing area.

\subsubsection{Jaboci field conjugate points}
\begin{figure}[h]
 \centering
 \subfloat[]{{\includegraphics[width=7cm]{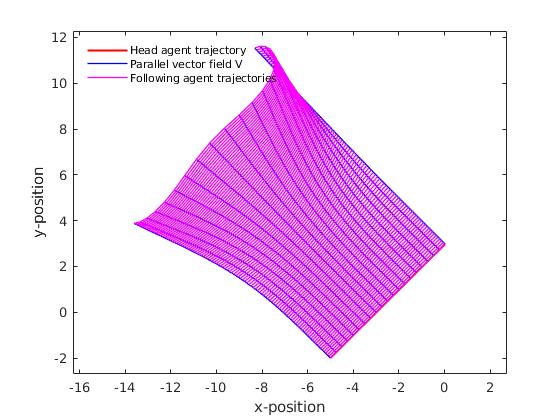}  }}%
 \\
 \subfloat[]{{\includegraphics[width=7cm]{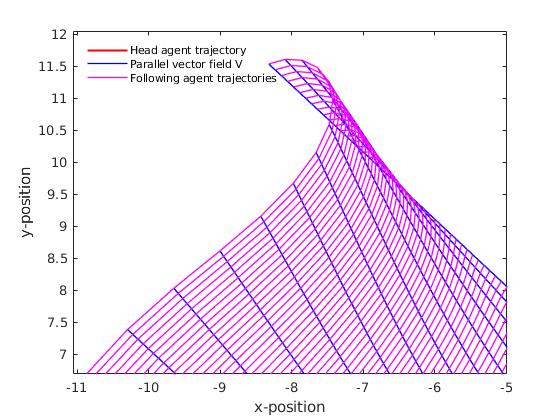} }}%
 \caption{Visual example of conjugate point. External potential $F(x_1, x_2)= \sin(\frac{x_1}{a}) + \cos(\frac{x_2}{a})$, $a=2$, head agent initial condition $[-5, -2, \cos(4/\pi), \sin(4/\pi) ]$. (a) Two dimensional view of swarm trajectory. Head agent trajectory in red, following agents' trajectory in magenta. (b) Enlargment of (a) around conjugate point.}
 \label{2dt_case4_i2_a4_t2_enlarge}
\end{figure}

 For external potential manifold with non-negative sectional curvature at each points, such as the elliptic paraboloid potential, there exist conjugate points in vector field $K$. Consider $p, q \in M$ are two points connected by a geodesic $\gamma$. $p, q$ are conjugate points along $\gamma$ if there exists a non-zero Jacobi field along $\gamma$ that vanishes at $p$ and $q$. Conjugate points are when the geodesic fails, locally, to be the minimum length curve connecting two points. Thus, our geodesic deviation based algorithm also fails. An visual example of conjugate point is shown in Fig. \ref{2dt_case4_i2_a4_t2_enlarge}. Additional agent's protocols needs to be installed to identify and bypass conjugate points.

\subsubsection{Striction curve}
\begin{figure}[h]
\centering
\subfloat[]{{\includegraphics[width=7cm]{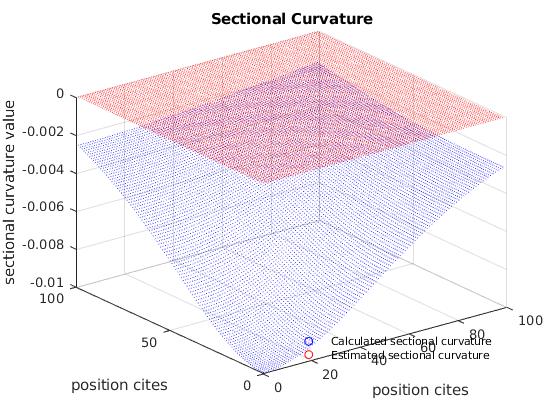}  }}%
\\
\subfloat[]{{\includegraphics[width=7cm]{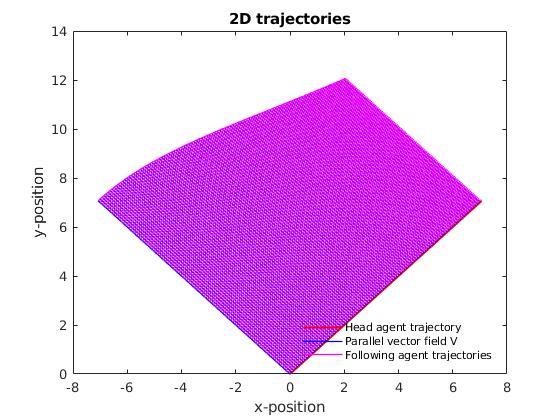} }}%

 \caption{Sectional curvature estimation and two-dimensional swarm trajectorise. External potential $F(x_1, x_2)=\frac{x_1^2-x_2^2}{a}$, $a=20$. Head agent travling on striction curve $[x_1, x_2]$. (a) Sectional curvature estimation. Estimated curvature in red; calculated curvature in blue. (b) Two dimensional view of swarm trajectories. Head agent trajectory in red, following agents' trajectory in magenta.}
 \label{feature2_explained}
\end{figure}

External potential two-dimensinal manifold with non-positive sectional curvature, such as the hyperbolic paraboloid, are saddle surface, and thus a ruled surface. For noncylindrical ruled surface, it always has a parameterization of the form
\begin{align}
 \mathbf{r}(u,v)= \boldsymbol{\sigma}(u)+v\boldsymbol{\delta}(u),
\end{align}
where $\boldsymbol{\sigma}$ is called the striction curve. In particular, hyperbolic paraboloid is a doubly ruled surface that has two striction curves. In our simulation of hyperbolic paraboloid potential, the two striction curves are $[x_1, \pm \, x_2]$. Shown in Fig. \ref{feature2_explained}, if the head agent is travling on striction curves, the swarm is unable to estimate the sectional curvature. The acceleration of separation vector field is zero, equivalent to a flat space (zero curvature).

With the same number of agents in the swarm, the communication frequency and inter-agent distance affects the sensing area. For external potential manifold with feature one, smaller sensing area is more likely to avoid conjugate points; while for external potential manifold with feature two, larger sensing area provides with more information about the manifold. The overall shape of the swarm reveals the external potential is not zero in Fig. \ref{feature2_explained}, eventhough the curvature estimation is zero.

\subsection{Formation control algorithm}
In the simulation for formation control algorithm in section \ref{Formation control}, we assume the agents can measure its relative distance to its neighbours. Head agent has its own course of trajectory, thus not affected by other agents in the swarm.
\begin{figure*}[t]
\centering
 \includegraphics[width=\textwidth]{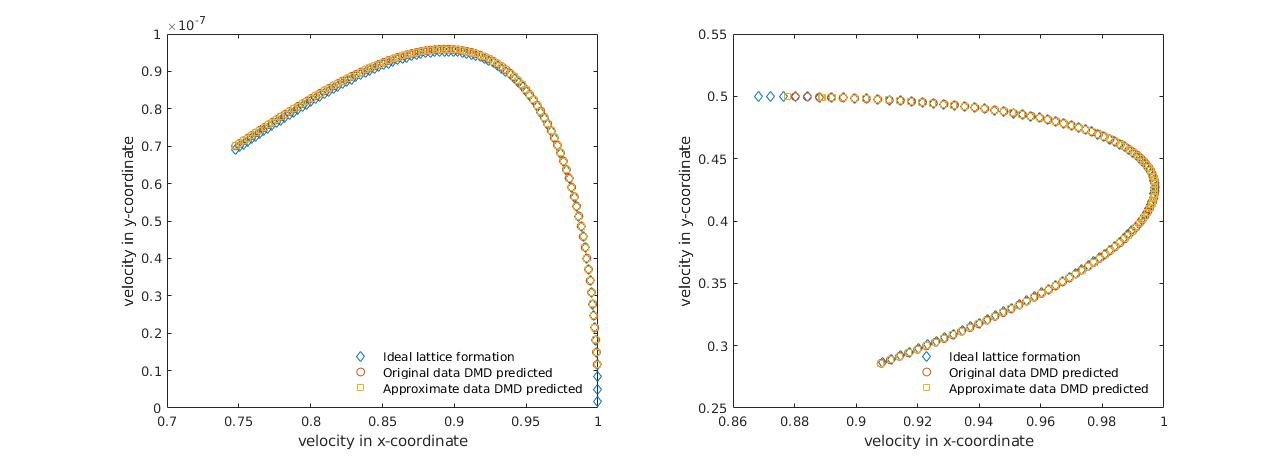}
 \caption{Velocity trajectory of the first following agent. Ideal lattice formation velocity in blue, original data DMD predicted velocity in red, approximate data DMD predicted velocity in yellow. External potential $F(x_1, x_2)=\displaystyle\frac{x_1^2+x_2^2}{a},\, a=20$. Head agent initial condition $[0, 0, 1, 0]$ (left), $[-5, 0, \cos(6/\pi), \sin(6/\pi)]$ (right).}
 \label{agent_velocity_DMD_compare_case2}
\end{figure*}
All following agent take its right-hand neighbour (closer to the head agent) as beacon to measure its own relative position. In ideal lattice formation, following agents are traveling in parallel and at a fixed distance to its right-hand neighbour, thus measurable in theory.

Fig. \ref{agent_velocity_DMD_compare_case2} shows the velocity trajectory of the following agent that is next to the head agent. In this simulation, we use timestep $\Delta t= 0.1$, data size $k=3$, observable $(\mathbf{x}_k,\mathbf{y}_k)$ to be the velocity of agent. In practice, the data $\mathbf{X}, \mathbf{Y}$ can only be collected by measuring the deviation of $k+1$ agents travling on nearby geodesics. Therefore, without any information of the external potential, the initial velocity of these agents can only be approximated in Eucliean metric, not the external potential metric itself. Fig. \ref{agent_velocity_DMD_compare_case2} shows that with reliable  correction step, the algorithm has reliable prediction of the agent's dynamic even uses data that is approximated in Eucliean metric.

However, when agents have near linear dynamic (constant velocity), the algorithm decreases in accuracy and eventually fails. This is expected as the algorithm is based on linear regression method. Therefore, an additional control protocol needs to be implanted when agents have linear dynamic, and the agents should have autonomous decision on when to switch the dynamic control protocols.

\subsection{Margin of algorithm}
In formation analysis algorithm simulations we have discussed for same numbers of agents, by changing communication frequency and inter-agent distance constrait, how the sensing area of the swarm is beneficial in some cases while in other cases affects the accuracy of the swarm formation analysis. Moreover, for the same sensing area, the number of agents have a positive correlation to the accuracy of the formation analysis.
Size of the swarm also plays a role in swarm formation control. In formation control algorithm simulations, the control algorithm requires agents to be able to communicate regarding their relative distance and angle in terms of the external potential manifold metric. In practice, this types of sensing is extremely difficult. Approximation of this inter-agent distance can be made in Eucliean metric by on-board agent sensories. We notice when using the same number of agents while either increases the inter-agent distance or increases the communication frequency to enlarge sensing area, the Euclidean metric approximate increases its error.

In summury, smaller swarm size is more controlable but with lower accuracy in external potential estimation, and vice versa. This conflict is due to the fact for formation analysis, we utilize the nonlinearity in agents' trajectory to estimate an nonlinear property, namely the external potential manifold curvature; while in formation control, we relay on linear approximation in Eucliean metric to correct the agent's trajectory predictions. Therefore, the balance between number of agents, communication frequency, and inter-agent distance is crucial in optimalize our approach of energy efficient WSNs algorithm.

\section{Conclusion}
We introduced a new approach for designing energy efficient WSNs algorithms inspired by swarm intelligence. In this approach, we identify clustering WSN as swarm, and external potentials as manifolds. By observing the change of swarm lattice formation in external potential, we are able to estimate the curvature of the manifold, which are valuable information of the external potential. To maintain lattice formation in external potential, fomation control algorithm uses DMD to predict the optimal agent's velocity in time thus guiding the trajectory of following agents.

\begin{acknowledgments}
Y.W is supported by the Japanese Government MEXT Scholarship Program. Y.W thanks V. Putkaradze for advices and encouragement. T.H acknowledges I. Mezic and Y. Susuki for helpful discussions.
\end{acknowledgments}

\bibliography{reference}
\end{document}